\documentclass[amssymb,prb,twocolumn,superscriptaddress]{revtex4-1}

\usepackage{bm}
\usepackage{epsfig}
\usepackage{graphicx}
\usepackage{amsmath}
\usepackage{color}

\usepackage[%
colorlinks=true,
urlcolor=blue,
linkcolor=blue,
citecolor=blue
]{hyperref}

\newcommand{\bra}[1]{{\langle #1|}}
\newcommand{\ket}[1]{{|#1\rangle}}

\newcommand{\comment}[1]{}

\newcommand{\bk}{{\bf k}}
\newcommand{\creatpair}[1]{c_{#1\uparrow}^\dagger c_{-#1\downarrow}^\dagger}

\begin{document}

\title{Topological $s$-wave superconductors driven by electron correlation}

\author{Huai-Shuang Zhu}
\affiliation{Department of Physics, Renmin University of China,
Beijing 100872, China}

\author{Zhidan Li}
\affiliation{Institute of Physics, Chinese Academy of Sciences,
Beijing, China}

\author{Qiang Han}
\email{hanqiang@ruc.edu.cn}
\affiliation{Department of Physics, Renmin University of China,
Beijing 100872, China}
\affiliation{Department of Physics and HKU-UCAS Joint Institute for Theoretical and Computational Physics at Hong Kong, The University of Hong Kong, Pokfulam Road,
Hong Kong, China}

\author{Z. D. Wang}
\email{zwang@hku.hk}
\affiliation{Department of Physics and HKU-UCAS Joint Institute for Theoretical and Computational Physics at Hong Kong, The University of Hong Kong, Pokfulam Road,
Hong Kong, China}
\date{\today}
\begin{abstract}
It is interesting to ask whether electron interaction can turn a topologically trivial superconductor into a nontrivial one without the presence of spin-obital coupling. In this paper we solve a correlated $s$-wave superconducting model exactly. The variation of the fermion number parity of the superconducting ground state as a function of  the electron interaction is calculated and the topological phase diagram is obtained. Topological $s$-wave superconducting states are revealed in the doped Mott insulators,
which is further confirmed by the numerical investigation of the topological boundary zero mode.
\end{abstract}

\maketitle

{\it Introduction.---} The last decade has witnessed the rapid progress in the research of topological phases of condensed matter, \cite{Hasan2010,Qi2011,Yan2017,Zhao2013,Hasan_2015,Zhao2016,Sato_2017,Armitage2018} including topological insulators, semimetals and superconductors. Among them, the topological superconductors have attracted tremendous interest for their potential application to realize fault tolerant quantum computation \cite{Nayak08}. The fully gaped superconducting ground states due to either intrinsic or effective spin-triplet $p$-wave pairing are believed to be the central ingredient in various theoretical proposals to realize topologically nontrivial superconductors. 
By taking advantage of the topological band structures of parent compounds, intrinsic topological superconductivity \cite{Mackenzie2003} are suggested to emerge in the bulk of the doped topological insulators \cite{Hor2010,Wray2010,Sato2013,Nikitin2016}.
There are also many proposals to establish effective $p$-wave pairing on the interface of heterostructures combining topological insulators \cite{FuL} or spin-orbital coupled semiconductors with the $s$,\cite{Sau2010,Lutchyn, Oreg} or $d$-wave \cite{Sato2010,Wong2012,Li2015} superconductors. The linear dispersive
low-lying electronic bands possessing spin-momentum locked internal structures are essential in the designs of realizing the bulk or interfacial topological superconductivity. 

The effects of electron-electron interaction on the topological superconducting phases have also been studied extensively.
Strong interactions were found to suppress topological superconducting phases by destroying the pairing-induced superconducting gap.~\cite{Gangadharaiah2011,Stoudenmire2011} Furthermore, phase transitions from topological superconducting phases to trivial conventional ones ~\cite{Fidkowski2010,Thomale2013,Katsura2015,Miao2017} were examined. In short, interaction can influence the stability of the topological phases by closing the bulk gap or inducing conventional competing orders due to spontaneous symmetry breaking. Despite the progress in the studies of the interaction effects, more fascinating effects of electron correlation are expected in the interacting topological superconductors. It is important to ask whether certain topologically nontrivial superconducting phases can be generated due to strong electron correlations, which do not have any non-interacting analogue. Indeed as in the insulator cases,~\cite{Raghu2008,Hohenadler2012,Morimoto2016,Rachel_2018,Kubo2019} several exotic states of matter such as topological Mott insulators~\cite{Pesin2010,Rachel2010,Wang2016} and fractional Chern insulators~\cite{Maciejko2015} have already been revealed, whose ground states do not adiabatically connect to those of the non-interacting topological insulators.

Motivated by the rich phenomena emerged in the interplay among superconductivity, correlations, and topologies, we study in this paper how a topologically trivial superconductor is driven into a nontrivial one in the presence of electron interactions. To realize this idea, a correlated electron model, the Hatsugai-Kohmoto (HK) model~\cite{HK}, is adopted where electron repulsion is local in 
momentum and the Mott physics can be captured exactly. Recently Phillips {\it et. al} ~\cite{PYH} demonstrated an analogue of the Cooper instability for the HK model of a doped Mott insulator. Comparing with their work~\cite{PYH} which studied exact superconducting instability in a non-Fermi liquid metallic state, here we focus on the topological characteristics of the $s$-wave superconducting states in the HK model. A fascinating observation in this paper lies in that a topologically nontrival $s$-wave superconducting ground state can be established by strong electron correlations without invoking spin-orbital coupling.

{\it Model and exact solution.---} The model Hamiltonian combines the HK model with the $s$-wave pairing, which is written as,
\begin{equation} \label{eqn:modelham}
	H = \sum_{\bk,\sigma} \xi_\bk \hat{n}_{\bk,\sigma} + \sum_\bk [ U \hat{n}_{\bk\uparrow}\hat{n}_{\bk\downarrow} + 
\Delta (c_{\bk\uparrow}c_{-\bk\downarrow} + \text{h.c.})]
\end{equation}
where $\xi_\bk=\epsilon_\bk - \mu$ represents the single-particle energy dispersion relative to the chemical potential $\mu$ and the electron number operator $\hat{n}_{\bk,\sigma} = c_{\bk\sigma}^\dagger c_{\bk\sigma}$. As a main spirit of HK model, $U$ signifies the electron interaction local in the momentum rather than the real space, which reflects the Mott physics in a more tractable manner. $\Delta $ denotes the conventional $s$-wave pairing potential which either arises from the proximity effect when the doped Mott insulator is adjacent to an $s$-wave superconductor or associates with Cooper pairing in the presence of the pairing interaction as demonstrated in Ref.~\cite{PYH}. 

\begin{figure}[ht]
 	\includegraphics[width=0.9\columnwidth]{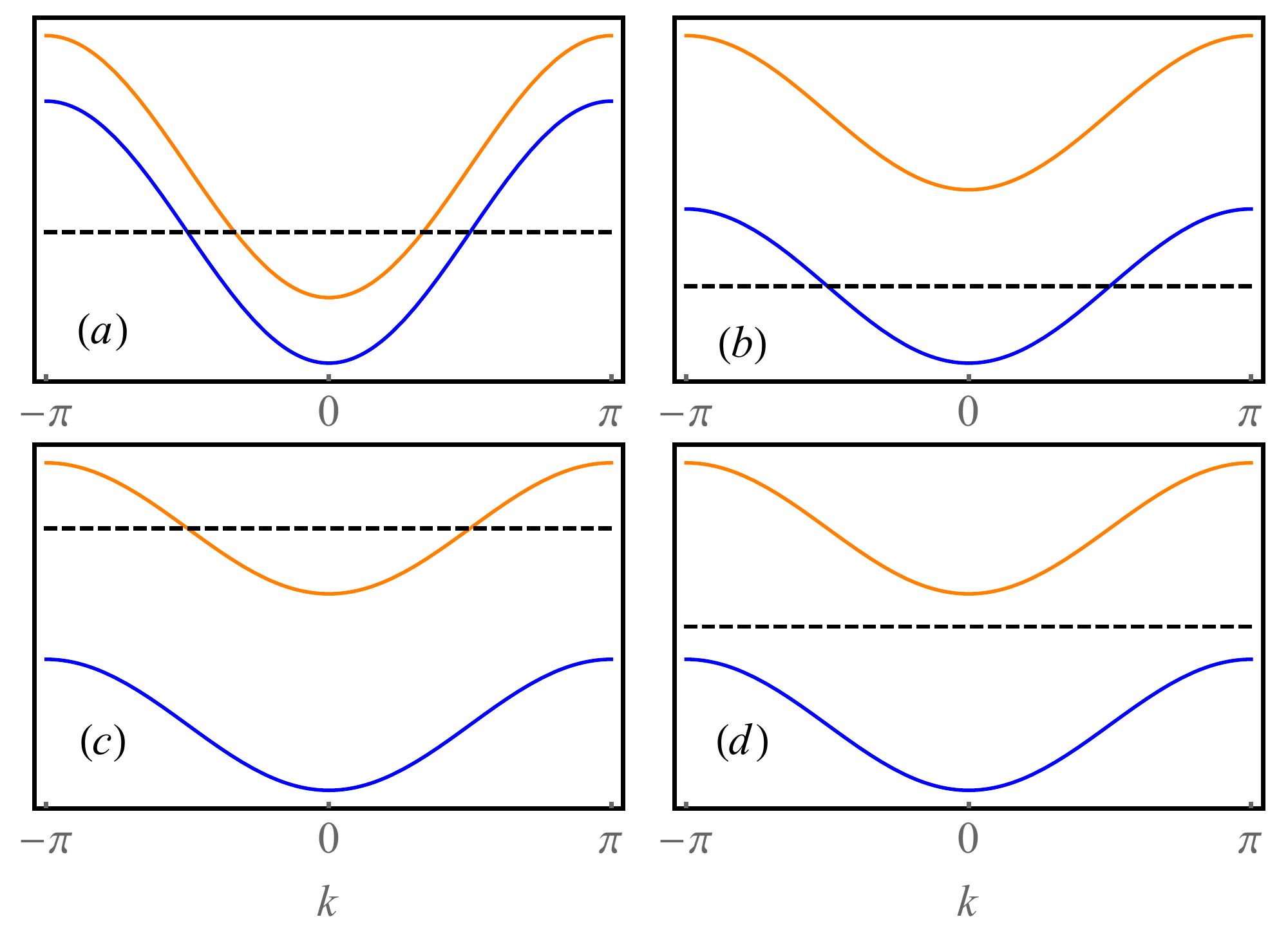}
 	\caption{Four normal phases of the HK model. Lower (LHB) and upper Hubbard bands (UHB), with energy dispersions $\xi_\bk$ and $\xi_\bk+U$ respectively, are plotted as blue and yellow lines. Here we let $\xi_\bk=-2t\cos k - \mu$ for illustration. Black dashed line denotes the position of chemical potential. (a), (b) and (c) represent three metallic phases with (a) both bands are partially occupied; (b) partically-occupied LHB and empty UHB; and
(c) fully-occupied LHB and partically-occupied UHB. (d) denotes the Mott insulator phase. }
 	\label{fig:HK}
\end{figure}

The HK model, i.e. the case of $\Delta=0$ in Eq.~(\ref{eqn:modelham}), is integrable and exactly solvable because $[H, \hat{n}_{\bk\sigma}]=0$ and $[\hat{n}_{\bk\sigma},\hat{n}_{\bk'\sigma'}]=0$ for any $\bk$ and $\bk'$, and accordingly each eigenstate of the HK model can be written as $\prod_\bk \ket{n_{\bk\uparrow}n_{\bk\downarrow}}$, where $n_{\bk\sigma}=0,1$ the
quauntum number of $\hat{n}_{\bk\sigma}$. In each $\bk$ sector, there are four eigenstates: $\ket{00}$, $\ket{10}$, $\ket{01}$, and $\ket{11}$, namely the empty, singly (spin up, down) and doubly occupied states, respectively. The corresponding eigenenergies are 0, $\xi_\bk$, and $2\xi_\bk+U$. From this knowledge, the many-body ground states of the HK model can be
readily obtained, whose four typical phases are shown in Fig.~\ref{fig:HK} for illustration. Note that the metallic phases are highly degenerate for the sake of the spin up and down degeneracy of the singly occupied band.

For $U=0$, the model is reduced to the mean-field BCS Hamiltonian $H_\text{BCS}=\sum_\bk H_\bk$  with
\begin{equation}
	H_\bk = \xi_\bk(\hat{n}_{\bk\uparrow}+\hat{n}_{-\bk\downarrow})+\Delta(c_{\bk\uparrow}c_{-\bk\downarrow}+h.c.).
\end{equation}
Since $H_\bk$'s are mutually commutative, the ground state of $H_\text{BCS}$ are therefore the product of that of each $H_\bk$. After the Bogoliubov quasiparticle transformation, the ground state of $H$ can then be written as
\begin{equation}
	\ket{\text{gs}} = \prod_\bk (u_\bk + v_\bk \creatpair{\bk})\ket{0},
\end{equation}
where $u_\bk$ and $v_\bk$ the coherent factors. The BCS ground state is obviously of even fermion parity and topologically trivial.

We then give the exact solution of Eq.~(\ref{eqn:modelham}) and investigate its topological properties by examining the interplay between the $U$ and $\Delta$ terms. Apparently the $U$ term couples the spin-up and down electrons with the same momentum, while the $\Delta$ terms couple the time-reversal related partners, which complicate the solution of $H$. However, the model Hamiltonian Eq.~(\ref{eqn:modelham}) is still tractable after decomposing it into the following form,
\begin{equation}
	H = \sum_\bk {}^\prime H_\bk^U,
\end{equation} 
where the $^\prime$  denotes that the summation over $\bk$ is restricted to half of the first Brillouin zone (HFBZ), e.g. $0\le k\le \pi$ in the 1D case. For $\bk$ in the interior of HFBZ (e.g. $0<k<\pi$ in 1D), $H_\bk^U$ is
\begin{equation} \label{ham:k}
	 H_\bk^U = H_\bk + H_{-\bk} + U (\hat{n}_{\bk\uparrow}\hat{n}_{\bk\downarrow}+\hat{n}_{-\bk\uparrow}\hat{n}_{-\bk\downarrow}),
\end{equation}
while for the momenta on the boundary of HFBZ ($k=0, \pi$ in 1D), i.e. the time-reversal-invariant momenta (TRIM) $\bk^*$,
\begin{equation}
	\label{ham:kstar}
	 H_{\bk^*}^U = H_{{\bk^*}} + U \hat{n}_{{\bk^*}\uparrow}\hat{n}_{{\bk^*}\downarrow}.
\end{equation}
All $H_{\bk(\bk^*)}^U$'s are commutative with each other and therefore the ground state of $H$ is the product of that
of each $H_{\bk(\bk^*)}^U$.

For any interior $\bk$ of HFBZ, the Fock space where $H_{\bk}^U$ lives is 16 dimensional, whose basis vectors are 
$\ket{n_1n_2n_3n_4}_\bk=c_{\bk\uparrow}^{\dagger{n_1}} c_{-\bk\downarrow}^{\dagger{n_2}} c_{-\bk\uparrow}^{\dagger{n_3}} c_{\bk\downarrow}^{\dagger{n_4}}\ket{0}_\bk$ with $n_i=0,1$. Since $H_\bk^U$ converves the fermion parity, this 16D Fock space can be further divided into even and odd sectors each of which is 8 dimensional.
In the even sector we find five doubly occupied eigenstates. Three of them are the spin triplet, i.e. $\ket{1010}_\bk$, $\frac{1}{\sqrt{2}}(\ket{1100}_\bk-\ket{0011}_\bk)$ and $\ket{0101}_\bk$. The other two are the Pauli blocking states, namely $\ket{1001}_\bk$, $\ket{0110}_\bk$. The spin triplet states obviously are immune to the $s$-wave pairing potential as well as the $U$ term, as reflected in their eigenenergy $2\xi_\bk$. The two Pauli blocking states do not take part in the pairing because they forbit Cooper pairs hopping into them according to the Pauli exclusion principle, and their energies are $2\xi_\bk+U$.
The remaining three eigenstates of $H_{\bk}^U$ in the even sector seat in the subspace spanned by $\{\ket{0000}_\bk, \frac{1}{\sqrt{2}}(\ket{1100}_\bk+\ket{0011}_\bk), \ket{1111}_\bk\}$, i.e. the empty, spin singlet and fully
occupied states. The matrix of $H_{\bk}^U$ in this 3D subspace is
\begin{equation} \label{subham}
	\begin{pmatrix}
		0 & -\sqrt{2}\Delta & 0\\
		-\sqrt{2}\Delta & 2\xi_\bk & -\sqrt{2}\Delta \\
		0 & -\sqrt{2}\Delta & 4\xi_\bk  + 2U
	\end{pmatrix}.
\end{equation}
Such a $3\times3$ symmetric matrix can be diagonalized analytically,~\cite{viete} and the three eigenenergies in ascending order are
\begin{equation}
	\lambda_{\bk}^{(i)} = 2\xi_\bk +\frac{2U}{3} + \frac{4}{\sqrt{3}} E_\bk^\text{even} \cos \left(\theta_\bk + \frac{2\pi}{3}i \right) \label{three:roots}
\end{equation}
where $i=1, 2, 3$,
\begin{align}
	& E_\bk^\text{even} =\sqrt{\big(\xi_\bk+\frac{U}{2}\big)^2+\Delta^2+\frac{U^2}{12}}, \label{Ek} \\
 & \theta_\bk  =\frac{1}{3} \arccos\big[\frac{q_\bk}{(\sqrt{3}E_\bk^\text{even})^3}\big], \label{thetak} \\
	& q_\bk = U\big( U^2 + \frac{9}{2}U\xi_\bk + \frac{9}{2} \xi_\bk^2 - \frac{9}{4}\Delta^2  \big). \label{qk}
\end{align} 
It can readily be found that $\lambda_\bk^{(1)}$ is the lowest-lying one among the eight eigenvalues regardless of
the variations of system parameters.

Similarly the 8D odd sector can be decomposed into 4 subspaces each of which is spanned by 2 basis vectors, e.g. $\ket{1000}_\bk$ and $\ket{1011}_\bk$. In each subspace, $H_\bk^U$ has the same $2\times2$ matrix form as
\begin{equation} 
	\begin{pmatrix}
		\xi_\bk & -\Delta \\
		-\Delta & 3\xi_\bk + U
	\end{pmatrix}
\end{equation}
The two eigenenergies are
\begin{equation} \label{two:roots}
	\lambda_{\bk\pm} = 2\xi_\bk + \frac{U}{2} \pm E_\bk^\text{odd}
\end{equation}
where 
\begin{equation}
	E_\bk^\text{odd}=\sqrt{\big(\xi_\bk+\frac{U}{2}\big)^2+\Delta^2}.
\end{equation}

Comparing the minumum energies in both sectors, we find that $\lambda_\bk^{(1)}$ is always lower than $\lambda_{\bk-}$. Since the eigenvector corresponding to $\lambda_\bk^{(1)}$ is in the even sector, the lowest-lying state of $H_\bk^U$ is definitely even in fermion parity. Thus the parity of the ground state of the whole system $H$ is solely determined by that of $H_{\bk^*}^U$. 

{\it The topological invariant and phase diagram.---} Comparing Eq.~(\ref{ham:kstar}) with Eq.~(\ref{ham:k}) we see that the Fock space of $H_{{\bk^*}}^U$ is 4D rather than 16D. This 4D Fock space is the direct sum of one even and one odd
subspaces, each of which is two dimensional. The even sector is spanned by $\ket{00}_{\bk^*}$ and $\ket{11}_{\bk^*}$, in which the Hamiltonian matrix is
\begin{equation}
	\begin{pmatrix}
		0 & -\Delta \\
		-\Delta & 2\xi_{\bk^*} + U \\
	\end{pmatrix}.
\end{equation}
In the even sector the lowest eigenenergy  of $H_{\bk^*}^U$ is $\xi_{\bk^*}+U/2-E_{\bk^*}^\text{odd}$, which is negative and approaches to
zero when $U\to\infty$.
In the odd sector, the spin doublet $\ket{10}_{\bk^*}$ and $\ket{01}_{\bk^*}$ are two degenerate eigenstates
of $H_{\bk^*}^U$ whose eigenvalue is $\xi_{\bk^*}$ independent of $U$. With increasing $U$, an energy level crossing will take place at a critical point $U_c =2E_{\bk^*}^\text{odd}$, accompanied by the parity switch from even to odd. The fermion parities of
all $H_{\bk^*}^U$'s together determine the fermion parity $\mathcal{P}$ of the ground state of whole system $H$. The fermion parity operator is defined as $\hat{P}=(-1)^{\sum_{\bk\sigma}\hat{n}_{\bk\sigma}}$, from which we obtain
\begin{align}
	\mathcal{P} & = \prod_{\bk^*} \text{sgn}(2E_{\bk^*}^\text{odd}-U), \label{P:1}\\
			 & = \prod_{\bk^*} \text{sgn}[ \xi_{\bk^*} (\xi_{\bk^*}+U)+\Delta^2 ] \label{P:2}
\end{align}
which acts as the topological invariant of this correlated $s$-wave superconducting model. 

From Eqs.~(\ref{P:1}) and (\ref{P:2})  Because $\mathcal{P}^2=1$, $\mathcal{P}$ is a many-body $\mathbb{Z}_2$ topological invariant. The uncorrelated $s$-wave superconductor with $U=0$ has $\mathcal{P}=1$ corresponding to the topologically trivial superconductor, while the ground state of nontrivial $s$-wave superconductor is of odd fermion parity with $\mathcal{P}=-1$. 
For weak pairing potential satisfying $\Delta\ll\min\{\xi_{\bk^*},\xi_{\bk^*}+U\}$, 
the topological index can be written approximately as $\mathcal{P}=\mathcal{P}_{\text{L}}\mathcal{P}_{\text{U}}$ with $\mathcal{P}_{\text{L}}=\prod_{\bk^*}\text{sgn}\xi_{\bk^*}$ and $\mathcal{P}_{\text{U}}=\prod_{\bk^*}\text{sgn}(\xi_{\bk^*}+U)$ denoting the fermion parities of the LHB and UHB, respectively.
Therefore, the position of the chemical potential $\mu$ with respective to the LHB and UHB, together with the lifting of the UHB above the LHB controlled by $U$, determines the topological nature of the ground states. For illustration, $\mathcal{P}_\text{L}=\text{sgn}(\xi_0 \xi_\pi)$ and $\mathcal{P}_\text{U}=\text{sgn}[(\xi_0+U)(\xi_\pi+U)]$ for the 1D case. When $s$-wave pairing is established in the weakly-correlated metallic phase as shown in Fig.~\ref{fig:HK}(a), both the LHB and the UHB are partially occupied and thus $\mathcal{P}_\text{L}=\mathcal{P}_\text{U}=-1$ resulting in the trivial conventional superconducting state. 
Enhencing the electron correlation by increasing $U$, a strongly-correlated metallic phase is reached with partially-occupied LHB and empty UHB corresponding to Fig.~\ref{fig:HK}(b). For this case we have  
$\mathcal{P}_\text{L}=-1$ and $\mathcal{P}_\text{U}=1$ and the resulting superconducting ground state is topologically nontrivial. Similar result can also be obtained for the metallic state shown in Fig.~\ref{fig:HK}(c).

\begin{figure}[ht]
 	\includegraphics[width=0.7\columnwidth]{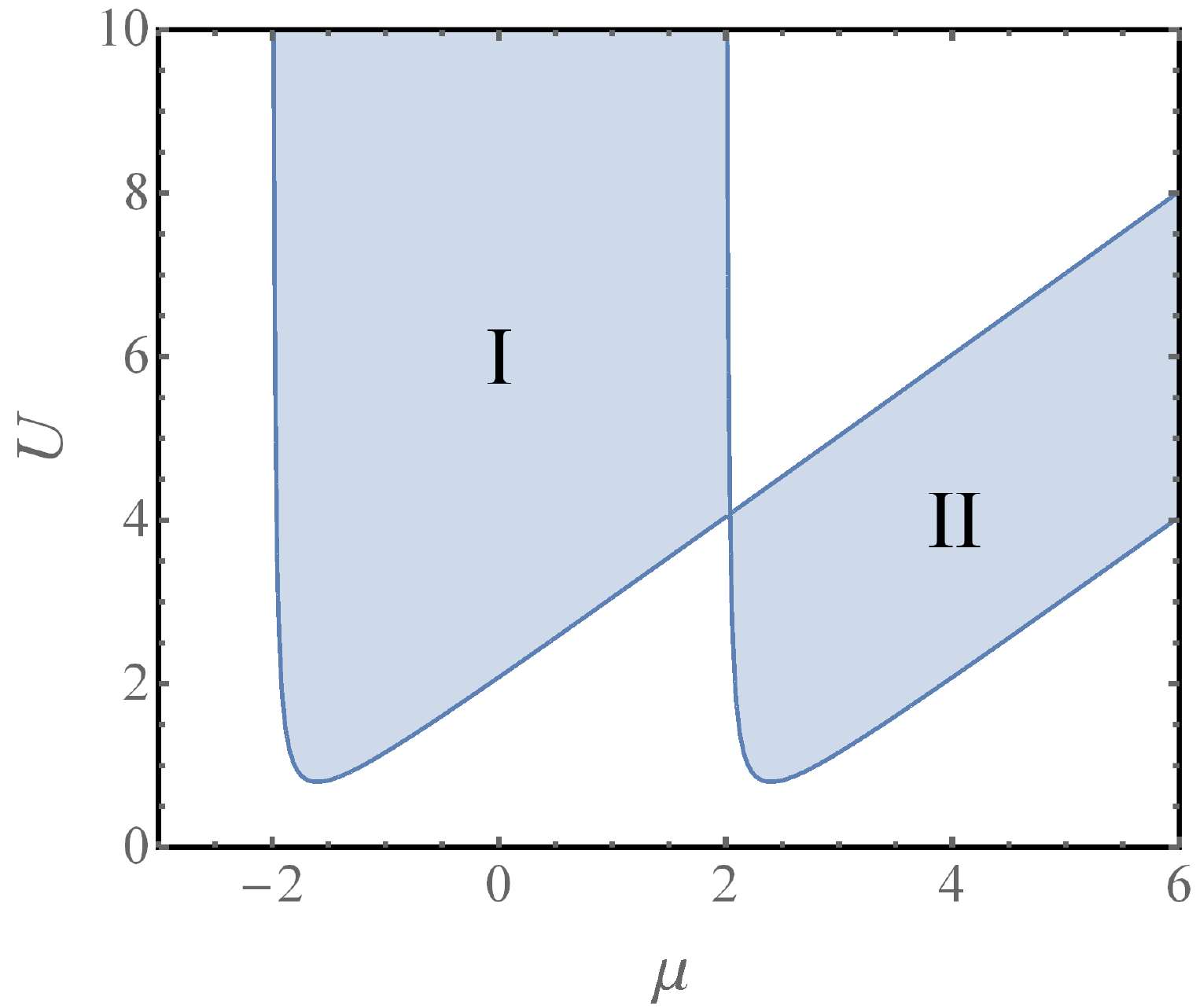}
 	\caption{Topological phase diagram of the 1D superconducting HK model. The single-electron energy dispersion is $\xi_k=-2t\cos k -\mu$. $t=1$ and the pairing potential $\Delta=0.4$.}
 	\label{fig:phasediagram}
\end{figure}

Figure \ref{fig:phasediagram} displays the topological phase diagram of our model system, where the phase boundaries separating the two topologically distinct phases are governed by
$
	\prod_{\bk^*} [ \xi_{\bk^*} (\xi_{\bk^*}+U)+\Delta^2 ] = 0.
$
The topologically nontrivial region are composed of two shaded areas denoted by I and II which correspond to
topological superconducting phases on the basis of the normal states (b) and (c) in Fig.~\ref{fig:HK}.

The above results show how the ground-state fermion parity as a many body topological invariant of superconductors~\cite{Shiozaki2016} varies with the electron correlations. This can also be understood from another perspective by examining how the ground state degeneracy is altered by the $s$-wave pairing. The ground states of all the three metallic phases as show in Fig.~\ref{fig:HK}(a)-(c) are highly degenerate. However, in the presence an $s$-wave pairing potential, the degeneracy of phase (a) is totally lifted. On the other hand, the two-fold degeneracy at certain TRIM is kept in the metallic phases (Fig.~\ref{fig:HK}(b) and (c)) because electron pairing can not compete with the strong correlation at the $\bk^*$.
 
{\it Topological boundary mode.}--- The topological superconducting state can also be identified by examining the boundary zero modes associated with the nontrivial bulk topology. For this purpose, the real-space conrespondence to the momentum space Hamiltonian of Eq.~(\ref{eqn:modelham}) is studied, which is expressed as
\begin{align}
H =& -t \sum_{j=1}^{L-1} (c_{j\sigma}^\dagger c_{j+1\sigma} + \text{h.c.}) - \mu \sum_{j=1}^{L} c_{j\sigma}^\dagger c_{j\sigma} \notag \\
& +\Delta \sum_{j=1}^L (c_{j\uparrow}c_{j\downarrow} + \text{h.c.}) \label{ham:realspace} \\
& + U/L \sum_{j_1,j_2,j_3,j_4=1}^L \delta_{j_1+j_3,j_2+j_4}c_{j_1\uparrow}^\dagger c_{j_2\uparrow} c_{j_3\downarrow}^\dagger c_{j_4\downarrow} \notag	
\end{align}
Here we choose a 1D lattice for simplicity, $L$ denotes the lattice length and the open boundary condition is chosen to examine the topological end modes. Note that the locality of the $U$ term in momentum space as in Eq.~(\ref{eqn:modelham}) results in the non-local
long-range interaction in real-space lattice as in Eq.~(\ref{ham:realspace}). The real-space Hamiltonian is solved by exact diagonalization in the Fock space using basis vectors
$\ket{\cdots n_{j\uparrow}\cdots;\cdots n_{j\downarrow}\cdots}$. The dimension of the Fock space is $2^{2L}$. Although the total fermion number is not conserved due to the pairing term, the fermion parity and the $z$-component of the total spin $M_z=N_\uparrow-N_\downarrow$ are good quantum numbers, which can help to reduce the matrix dimension. Actually, the fermi parity and $M_z$ are closely related because $\mathcal{P}=(-1)^{N_\uparrow+N_\downarrow}=(-1)^{M_z}$. Thus $M_z$ is used to label each sector of the Fock space.  
Furthermore since the time reversal symmetry is reserved, the ground states in the $M_z=\pm1$ sectors are degenerate according to the Kramers' theorem. Therefore we only need to
perform numerical diagonalization  of Eq.~(\ref{ham:realspace}) in the $M_z=0,1$ sectors for the purpose of exploring the end modes.
	
\begin{figure}[ht]
 	\includegraphics[width=0.7\columnwidth]{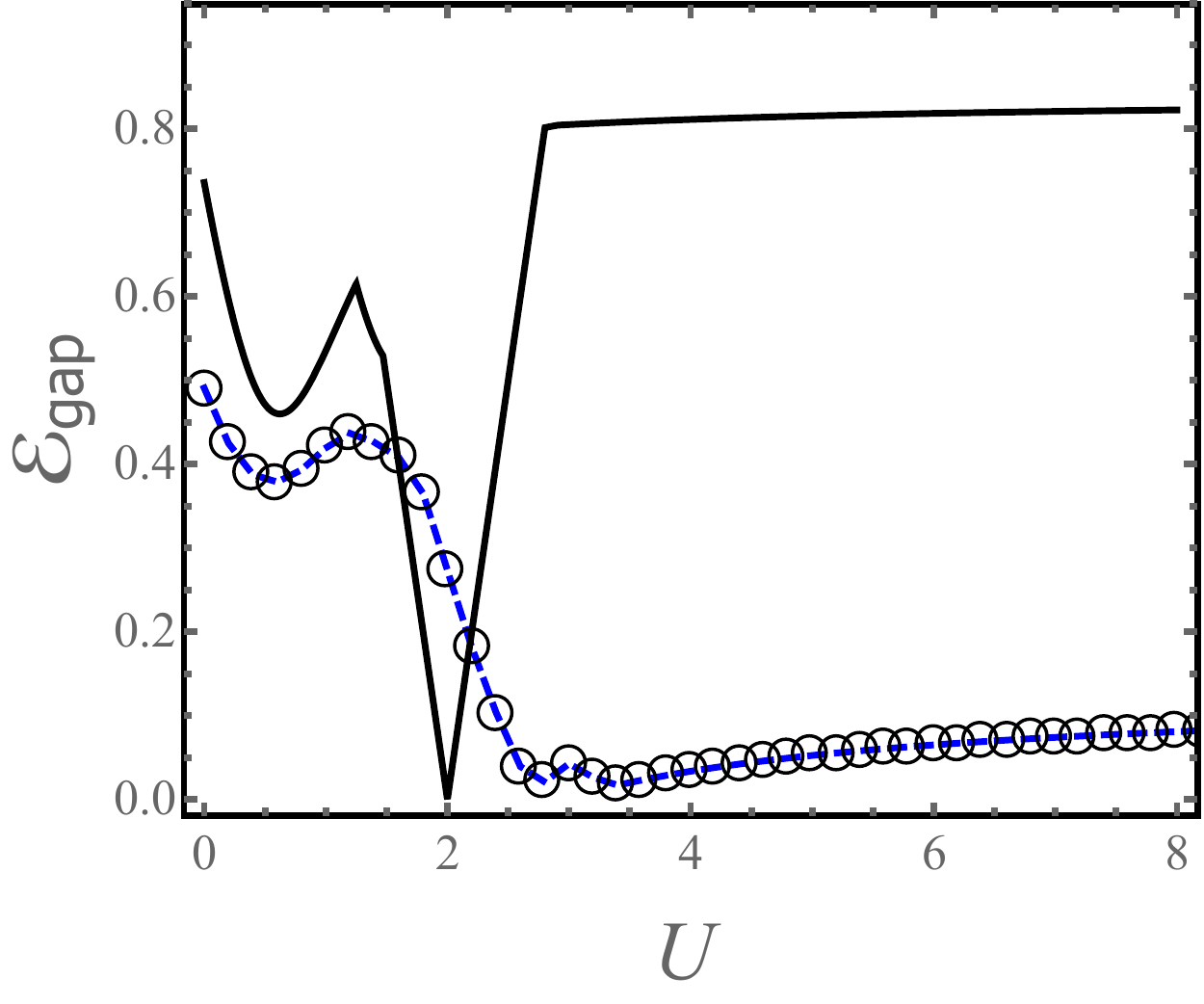}
 	\caption{Single-particle excitation gap $\varepsilon_\text{gap}$ as a function of $U$ in the 1D superconducting HK chain with $L=10$. $\varepsilon_\text{gap}$ is measured as the difference between $\varepsilon_\text{g}^\text{even}$ and $\varepsilon_\text{g}^\text{odd}$ which are the energies of the ground states in the even and odd sectors. The system parameters are $t=1$, $\mu=0$ and $\Delta=0.4$. Solid black line and open circles threaded by dashed blue line correspond to periodic and open boundary conditions, respectively. 
 }
\label{fig:gap}
\end{figure}
Fig.~\ref{fig:gap} illustrates the $U$ dependence of the single-particle excitation gap of our $s$-wave superconducting HK model for open (OBC) and periodic boundary conditions (PBC). The gap is calculated according to $\varepsilon_\text{gap}=|\varepsilon_\text{g}^{(0)}-\varepsilon_\text{g}^{(1)}|$, where $\varepsilon_\text{g}^{(M_z)}$ denotes the ground-state energy of the system in the $M_z$ sector. In our investigation, the model parameters are $t=1$ as the unit of energy, $\mu=0$ giving rise to half filling of the LHB as shown in Fig.~\ref{fig:HK}(b), and $\Delta=0.4$. In calculating the variation of the single-particle excitation gap as a function of $U$, a shorter chain with $L=10$ is chosen.  The PBC results, obtained from our analytic results Eqs.~(\ref{three:roots}) and (\ref{two:roots}), indicate that the gap first closes at and then reopen above $U_c=2.08$. For the OBC case, the gap closes around $U=2.8$ and remains nearly closed above it implying the existence of an approximate zero mode with $\varepsilon_\text{gap}\approx 0.04 \ll \Delta$ inside the superconducting gap. The deviation of the critical $U_c$'s obtained for PBC and OBC as well as the small but nonvanishing excitation energy are attributed to the finite size of the system.

\begin{figure}[ht]
 	\includegraphics[width=0.7\columnwidth]{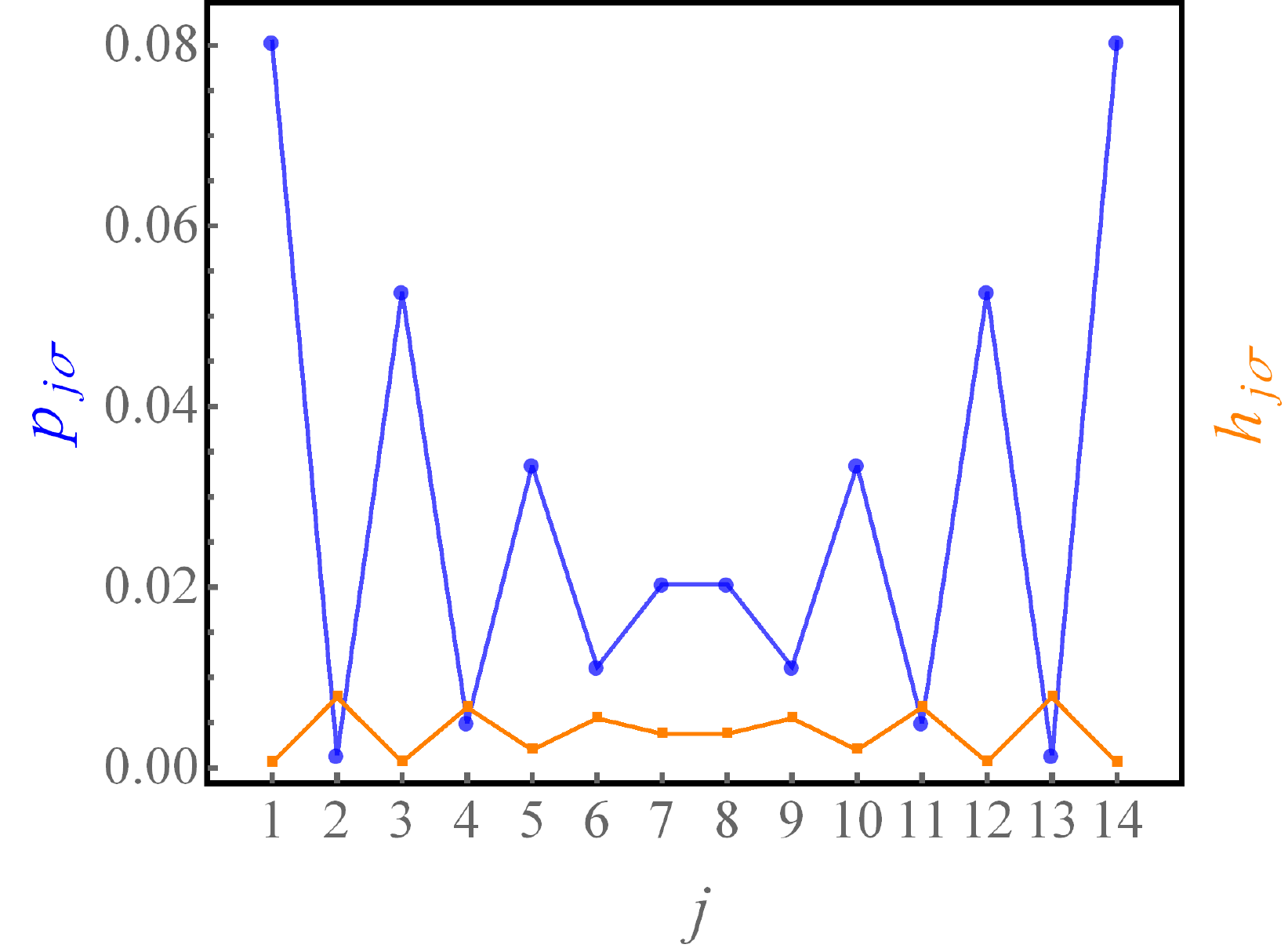}
 	\caption{Spatial distribution of the probability of adding a single electron ($p_{j\sigma}$) or a single hole ($h_{j\sigma}$) to the ground state. The chain length is $L=14$, with other parameters $\mu=0$, $\Delta=0.4$, and $U/L=2$ in the topological region I in Fig.~\ref{fig:phasediagram}
 }
\label{fig:ldos}
\end{figure}
To further check whether this zero mode is localized around the ends of the chain, we study the low-energy local density of states (LDOS), which can manifest the spatial distribution of the zero mode. The contribution of the zero mode to the LDOS can be extracted from the imaginary part of the single-particle retarded Green's function and written as,
\begin{equation}
\rho_{j\sigma}(\omega) = p_{j\sigma}
\delta(\omega-\varepsilon_{\text{g}}^{(1)}+\varepsilon_{\text{g}}^{(0)}) 
																		+  h_{j\sigma}\delta(\omega+\varepsilon_{\text{g}}^{(1)}-\varepsilon_{\text{g}}^{(0)}) 
\end{equation}
where the $p_{j\sigma}(h_{j\sigma})$ represents the probability of adding(removing) a spin-$\sigma$ electron on site $j$ of the chain. $p_{j\uparrow}$ and $h_{j\uparrow}$  are calculated according to
\begin{align}
& p_{j\uparrow} = \big| \bra{ \psi_\text{g}^{(1)} }c_{j\uparrow}^{\dagger} \ket{ \psi_\text{g}^{(0)} } \big|^2,  \\
& h_{j\uparrow} = \big| \bra{\psi_\text{g}^{(-1)}}c_{j\uparrow} \ket{ \psi_\text{g}^{(0)} } \big|^2
\end{align}
Here $\ket{ \psi_\text{g}^{(M_z)} }$ denotes the ground-state eigenvector of Eq.~(\ref{ham:realspace}) in the $M_z$ sector of the Fock space. By using the time-reversal symmetry we have $\ket{ \psi_\text{g}^{(-1)} }=\hat{T}\ket{ \psi_\text{g}^{(1)} }$ and $\hat{T}c_{j\uparrow}\hat{T}^{-1}=-c_{j\downarrow}$  with $\hat{T}$ the time reversal operator, and therefore $p_{j\uparrow}=p_{j\downarrow}$ and $h_{j\uparrow}=h_{j\downarrow}=\big| \bra{\psi_\text{g}^{(1)}}c_{j\downarrow} \ket{ \psi_\text{g}^{(0)} } \big|^2$. In the investigation of the boundary mode, the model parameters are the same as those in Fig.~\ref{fig:gap} except that we study a longer chain with $L=14$ and the interaction $U/L=2$ corresponding to a topological phase in Fig.~\ref{fig:phasediagram}. This results in the energy difference between ground states in the the even and odd sectors is as small as $0.004$, further supporting the emergence of the zero mode.
As for the spatial distribution of the zero mode, $p_{j\sigma}$ and $h_{j\sigma}$ as a function of $j$ are ploted in Fig.~\ref{fig:ldos}. Both $p_{j\sigma}$ and $h_{j\sigma}$ have greater values near the chain boundaries and decay fastly into the bulk, exhibiting localized distribution of the zero modes. Furthermore,  $p_{j\sigma}$ is about ten times larger on average than $h_{j\sigma}$ and this asymmetry between adding single particle and hole into the ground state is caused by the strong electron correlation. Therefore, the many-body boundary zero mode is a fermion-like zero mode rather than a Majorana one.

In summary, a correlated $s$-wave superconducting model whose normal states are doped Mott insulating phases is investigated.
This model is exactly solved taking advantage of the fact that the electron-electron interaction is local in the momentum space. 
The fermionic parity of the superconducting ground state, which acts as the many-body topological invariant, is derived explicitly and found to be related merely to the time-reversal-invariant momenta. The 1D lattice model with the open boundary condition is examined by the exact diagonalization, and the excitation energy and spatial distribution of the boundary mode  further identify the interaction-induced topological superconducting phase.

\section{acknowledgement}
This work was supported by
the Key-Area Research and Development Program of GuangDong Province (Grant No. 2019B030330001),
the National Key Research and Development Program of China (Grant No. 2016YFA0301800),
the GRF (No.: HKU173057/17P) and CRF (No.: C6005-17G) of Hong Kong.


\bibliography{arxiv}

\begin{thebibliography}{40}%
\makeatletter
\providecommand \@ifxundefined [1]{%
 \@ifx{#1\undefined}
}%
\providecommand \@ifnum [1]{%
 \ifnum #1\expandafter \@firstoftwo
 \else \expandafter \@secondoftwo
 \fi
}%
\providecommand \@ifx [1]{%
 \ifx #1\expandafter \@firstoftwo
 \else \expandafter \@secondoftwo
 \fi
}%
\providecommand \natexlab [1]{#1}%
\providecommand \enquote  [1]{``#1''}%
\providecommand \bibnamefont  [1]{#1}%
\providecommand \bibfnamefont [1]{#1}%
\providecommand \citenamefont [1]{#1}%
\providecommand \href@noop [0]{\@secondoftwo}%
\providecommand \href [0]{\begingroup \@sanitize@url \@href}%
\providecommand \@href[1]{\@@startlink{#1}\@@href}%
\providecommand \@@href[1]{\endgroup#1\@@endlink}%
\providecommand \@sanitize@url [0]{\catcode `\\12\catcode `\$12\catcode
  `\&12\catcode `\#12\catcode `\^12\catcode `\_12\catcode `\%12\relax}%
\providecommand \@@startlink[1]{}%
\providecommand \@@endlink[0]{}%
\providecommand \url  [0]{\begingroup\@sanitize@url \@url }%
\providecommand \@url [1]{\endgroup\@href {#1}{\urlprefix }}%
\providecommand \urlprefix  [0]{URL }%
\providecommand \Eprint [0]{\href }%
\providecommand \doibase [0]{http://dx.doi.org/}%
\providecommand \selectlanguage [0]{\@gobble}%
\providecommand \bibinfo  [0]{\@secondoftwo}%
\providecommand \bibfield  [0]{\@secondoftwo}%
\providecommand \translation [1]{[#1]}%
\providecommand \BibitemOpen [0]{}%
\providecommand \bibitemStop [0]{}%
\providecommand \bibitemNoStop [0]{.\EOS\space}%
\providecommand \EOS [0]{\spacefactor3000\relax}%
\providecommand \BibitemShut  [1]{\csname bibitem#1\endcsname}%
\let\auto@bib@innerbib\@empty
\bibitem [{\citenamefont {Hasan}\ and\ \citenamefont {Kane}(2010)}]{Hasan2010}%
  \BibitemOpen
  \bibfield  {author} {\bibinfo {author} {\bibfnamefont {M.~Z.}\ \bibnamefont
  {Hasan}}\ and\ \bibinfo {author} {\bibfnamefont {C.~L.}\ \bibnamefont
  {Kane}},\ }\href {\doibase 10.1103/RevModPhys.82.3045} {\bibfield  {journal}
  {\bibinfo  {journal} {Rev. Mod. Phys.}\ }\textbf {\bibinfo {volume} {82}},\
  \bibinfo {pages} {3045} (\bibinfo {year} {2010})}\BibitemShut {NoStop}%
\bibitem [{\citenamefont {Qi}\ and\ \citenamefont {Zhang}(2011)}]{Qi2011}%
  \BibitemOpen
  \bibfield  {author} {\bibinfo {author} {\bibfnamefont {X.-L.}\ \bibnamefont
  {Qi}}\ and\ \bibinfo {author} {\bibfnamefont {S.-C.}\ \bibnamefont {Zhang}},\
  }\href {\doibase 10.1103/RevModPhys.83.1057} {\bibfield  {journal} {\bibinfo
  {journal} {Rev. Mod. Phys.}\ }\textbf {\bibinfo {volume} {83}},\ \bibinfo
  {pages} {1057} (\bibinfo {year} {2011})}\BibitemShut {NoStop}%
\bibitem [{\citenamefont {Yan}\ and\ \citenamefont {Felser}(2017)}]{Yan2017}%
  \BibitemOpen
  \bibfield  {author} {\bibinfo {author} {\bibfnamefont {B.}~\bibnamefont
  {Yan}}\ and\ \bibinfo {author} {\bibfnamefont {C.}~\bibnamefont {Felser}},\
  }\href {\doibase 10.1146/annurev-conmatphys-031016-025458} {\bibfield
  {journal} {\bibinfo  {journal} {Annu. Rev. Condens. Matter Phys.}\ }\textbf
  {\bibinfo {volume} {8}},\ \bibinfo {pages} {337} (\bibinfo {year}
  {2017})}\BibitemShut {NoStop}%
\bibitem [{\citenamefont {Zhao}\ and\ \citenamefont {Wang}(2013)}]{Zhao2013}%
  \BibitemOpen
  \bibfield  {author} {\bibinfo {author} {\bibfnamefont {Y.~X.}\ \bibnamefont
  {Zhao}}\ and\ \bibinfo {author} {\bibfnamefont {Z.~D.}\ \bibnamefont
  {Wang}},\ }\href {\doibase 10.1103/PhysRevLett.110.240404} {\bibfield
  {journal} {\bibinfo  {journal} {Phys. Rev. Lett.}\ }\textbf {\bibinfo
  {volume} {110}},\ \bibinfo {pages} {240404} (\bibinfo {year}
  {2013})}\BibitemShut {NoStop}%
\bibitem [{\citenamefont {Hasan}\ \emph {et~al.}(2015)\citenamefont {Hasan},
  \citenamefont {Xu},\ and\ \citenamefont {Bian}}]{Hasan_2015}%
  \BibitemOpen
  \bibfield  {author} {\bibinfo {author} {\bibfnamefont {M.~Z.}\ \bibnamefont
  {Hasan}}, \bibinfo {author} {\bibfnamefont {S.-Y.}\ \bibnamefont {Xu}}, \
  and\ \bibinfo {author} {\bibfnamefont {G.}~\bibnamefont {Bian}},\ }\href
  {\doibase 10.1088/0031-8949/2015/t164/014001} {\bibfield  {journal} {\bibinfo
   {journal} {Phys. Scr.}\ }\textbf {\bibinfo {volume} {T164}},\ \bibinfo
  {pages} {014001} (\bibinfo {year} {2015})}\BibitemShut {NoStop}%
\bibitem [{\citenamefont {Zhao}\ \emph {et~al.}(2016)\citenamefont {Zhao},
  \citenamefont {Schnyder},\ and\ \citenamefont {Wang}}]{Zhao2016}%
  \BibitemOpen
  \bibfield  {author} {\bibinfo {author} {\bibfnamefont {Y.~X.}\ \bibnamefont
  {Zhao}}, \bibinfo {author} {\bibfnamefont {A.~P.}\ \bibnamefont {Schnyder}},
  \ and\ \bibinfo {author} {\bibfnamefont {Z.~D.}\ \bibnamefont {Wang}},\
  }\href {\doibase 10.1103/PhysRevLett.116.156402} {\bibfield  {journal}
  {\bibinfo  {journal} {Phys. Rev. Lett.}\ }\textbf {\bibinfo {volume} {116}},\
  \bibinfo {pages} {156402} (\bibinfo {year} {2016})}\BibitemShut {NoStop}%
\bibitem [{\citenamefont {Sato}\ and\ \citenamefont {Ando}(2017)}]{Sato_2017}%
  \BibitemOpen
  \bibfield  {author} {\bibinfo {author} {\bibfnamefont {M.}~\bibnamefont
  {Sato}}\ and\ \bibinfo {author} {\bibfnamefont {Y.}~\bibnamefont {Ando}},\
  }\href {\doibase 10.1088/1361-6633/aa6ac7} {\bibfield  {journal} {\bibinfo
  {journal} {Rep. Prog. Phys.}\ }\textbf {\bibinfo {volume} {80}},\ \bibinfo
  {pages} {076501} (\bibinfo {year} {2017})}\BibitemShut {NoStop}%
\bibitem [{\citenamefont {Armitage}\ \emph {et~al.}(2018)\citenamefont
  {Armitage}, \citenamefont {Mele},\ and\ \citenamefont
  {Vishwanath}}]{Armitage2018}%
  \BibitemOpen
  \bibfield  {author} {\bibinfo {author} {\bibfnamefont {N.~P.}\ \bibnamefont
  {Armitage}}, \bibinfo {author} {\bibfnamefont {E.~J.}\ \bibnamefont {Mele}},
  \ and\ \bibinfo {author} {\bibfnamefont {A.}~\bibnamefont {Vishwanath}},\
  }\href {\doibase 10.1103/RevModPhys.90.015001} {\bibfield  {journal}
  {\bibinfo  {journal} {Rev. Mod. Phys.}\ }\textbf {\bibinfo {volume} {90}},\
  \bibinfo {pages} {015001} (\bibinfo {year} {2018})}\BibitemShut {NoStop}%
\bibitem [{\citenamefont {Nayak}\ \emph {et~al.}(2008)\citenamefont {Nayak},
  \citenamefont {Simon}, \citenamefont {Stern}, \citenamefont {Freedman},\ and\
  \citenamefont {Das~Sarma}}]{Nayak08}%
  \BibitemOpen
  \bibfield  {author} {\bibinfo {author} {\bibfnamefont {C.}~\bibnamefont
  {Nayak}}, \bibinfo {author} {\bibfnamefont {S.~H.}\ \bibnamefont {Simon}},
  \bibinfo {author} {\bibfnamefont {A.}~\bibnamefont {Stern}}, \bibinfo
  {author} {\bibfnamefont {M.}~\bibnamefont {Freedman}}, \ and\ \bibinfo
  {author} {\bibfnamefont {S.}~\bibnamefont {Das~Sarma}},\ }\href {\doibase
  10.1103/RevModPhys.80.1083} {\bibfield  {journal} {\bibinfo  {journal} {Rev.
  Mod. Phys.}\ }\textbf {\bibinfo {volume} {80}},\ \bibinfo {pages} {1083}
  (\bibinfo {year} {2008})}\BibitemShut {NoStop}%
\bibitem [{\citenamefont {Mackenzie}\ and\ \citenamefont
  {Maeno}(2003)}]{Mackenzie2003}%
  \BibitemOpen
  \bibfield  {author} {\bibinfo {author} {\bibfnamefont {A.~P.}\ \bibnamefont
  {Mackenzie}}\ and\ \bibinfo {author} {\bibfnamefont {Y.}~\bibnamefont
  {Maeno}},\ }\href {\doibase 10.1103/RevModPhys.75.657} {\bibfield  {journal}
  {\bibinfo  {journal} {Rev. Mod. Phys.}\ }\textbf {\bibinfo {volume} {75}},\
  \bibinfo {pages} {657} (\bibinfo {year} {2003})}\BibitemShut {NoStop}%
\bibitem [{\citenamefont {Hor}\ \emph {et~al.}(2010)\citenamefont {Hor},
  \citenamefont {Williams}, \citenamefont {Checkelsky}, \citenamefont
  {Roushan}, \citenamefont {Seo}, \citenamefont {Xu}, \citenamefont
  {Zandbergen}, \citenamefont {Yazdani}, \citenamefont {Ong},\ and\
  \citenamefont {Cava}}]{Hor2010}%
  \BibitemOpen
  \bibfield  {author} {\bibinfo {author} {\bibfnamefont {Y.~S.}\ \bibnamefont
  {Hor}}, \bibinfo {author} {\bibfnamefont {A.~J.}\ \bibnamefont {Williams}},
  \bibinfo {author} {\bibfnamefont {J.~G.}\ \bibnamefont {Checkelsky}},
  \bibinfo {author} {\bibfnamefont {P.}~\bibnamefont {Roushan}}, \bibinfo
  {author} {\bibfnamefont {J.}~\bibnamefont {Seo}}, \bibinfo {author}
  {\bibfnamefont {Q.}~\bibnamefont {Xu}}, \bibinfo {author} {\bibfnamefont
  {H.~W.}\ \bibnamefont {Zandbergen}}, \bibinfo {author} {\bibfnamefont
  {A.}~\bibnamefont {Yazdani}}, \bibinfo {author} {\bibfnamefont {N.~P.}\
  \bibnamefont {Ong}}, \ and\ \bibinfo {author} {\bibfnamefont {R.~J.}\
  \bibnamefont {Cava}},\ }\href {\doibase 10.1103/PhysRevLett.104.057001}
  {\bibfield  {journal} {\bibinfo  {journal} {Phys. Rev. Lett.}\ }\textbf
  {\bibinfo {volume} {104}},\ \bibinfo {pages} {057001} (\bibinfo {year}
  {2010})}\BibitemShut {NoStop}%
\bibitem [{\citenamefont {Wray}\ \emph {et~al.}(2010)\citenamefont {Wray},
  \citenamefont {Xu}, \citenamefont {Xia}, \citenamefont {Hor}, \citenamefont
  {Qian}, \citenamefont {Fedorov}, \citenamefont {Lin}, \citenamefont {Bansil},
  \citenamefont {Cava},\ and\ \citenamefont {Hasan}}]{Wray2010}%
  \BibitemOpen
  \bibfield  {author} {\bibinfo {author} {\bibfnamefont {L.~A.}\ \bibnamefont
  {Wray}}, \bibinfo {author} {\bibfnamefont {S.-y.}\ \bibnamefont {Xu}},
  \bibinfo {author} {\bibfnamefont {Y.}~\bibnamefont {Xia}}, \bibinfo {author}
  {\bibfnamefont {Y.~S.}\ \bibnamefont {Hor}}, \bibinfo {author} {\bibfnamefont
  {D.}~\bibnamefont {Qian}}, \bibinfo {author} {\bibfnamefont {A.~V.}\
  \bibnamefont {Fedorov}}, \bibinfo {author} {\bibfnamefont {H.}~\bibnamefont
  {Lin}}, \bibinfo {author} {\bibfnamefont {A.}~\bibnamefont {Bansil}},
  \bibinfo {author} {\bibfnamefont {R.~J.}\ \bibnamefont {Cava}}, \ and\
  \bibinfo {author} {\bibfnamefont {M.~Z.}\ \bibnamefont {Hasan}},\ }\href
  {\doibase 10.1038/nphys1762} {\bibfield  {journal} {\bibinfo  {journal} {Nat.
  Phys.}\ }\textbf {\bibinfo {volume} {6}},\ \bibinfo {pages} {855} (\bibinfo
  {year} {2010})}\BibitemShut {NoStop}%
\bibitem [{\citenamefont {Sato}\ \emph {et~al.}(2013)\citenamefont {Sato},
  \citenamefont {Tanaka}, \citenamefont {Nakayama}, \citenamefont {Souma},
  \citenamefont {Takahashi}, \citenamefont {Sasaki}, \citenamefont {Ren},
  \citenamefont {Taskin}, \citenamefont {Segawa},\ and\ \citenamefont
  {Ando}}]{Sato2013}%
  \BibitemOpen
  \bibfield  {author} {\bibinfo {author} {\bibfnamefont {T.}~\bibnamefont
  {Sato}}, \bibinfo {author} {\bibfnamefont {Y.}~\bibnamefont {Tanaka}},
  \bibinfo {author} {\bibfnamefont {K.}~\bibnamefont {Nakayama}}, \bibinfo
  {author} {\bibfnamefont {S.}~\bibnamefont {Souma}}, \bibinfo {author}
  {\bibfnamefont {T.}~\bibnamefont {Takahashi}}, \bibinfo {author}
  {\bibfnamefont {S.}~\bibnamefont {Sasaki}}, \bibinfo {author} {\bibfnamefont
  {Z.}~\bibnamefont {Ren}}, \bibinfo {author} {\bibfnamefont {A.~A.}\
  \bibnamefont {Taskin}}, \bibinfo {author} {\bibfnamefont {K.}~\bibnamefont
  {Segawa}}, \ and\ \bibinfo {author} {\bibfnamefont {Y.}~\bibnamefont
  {Ando}},\ }\href {\doibase 10.1103/PhysRevLett.110.206804} {\bibfield
  {journal} {\bibinfo  {journal} {Phys. Rev. Lett.}\ }\textbf {\bibinfo
  {volume} {110}},\ \bibinfo {pages} {206804} (\bibinfo {year}
  {2013})}\BibitemShut {NoStop}%
\bibitem [{\citenamefont {Nikitin}\ \emph {et~al.}(2016)\citenamefont
  {Nikitin}, \citenamefont {Pan}, \citenamefont {Huang}, \citenamefont {Naka},\
  and\ \citenamefont {de~Visser}}]{Nikitin2016}%
  \BibitemOpen
  \bibfield  {author} {\bibinfo {author} {\bibfnamefont {A.~M.}\ \bibnamefont
  {Nikitin}}, \bibinfo {author} {\bibfnamefont {Y.}~\bibnamefont {Pan}},
  \bibinfo {author} {\bibfnamefont {Y.~K.}\ \bibnamefont {Huang}}, \bibinfo
  {author} {\bibfnamefont {T.}~\bibnamefont {Naka}}, \ and\ \bibinfo {author}
  {\bibfnamefont {A.}~\bibnamefont {de~Visser}},\ }\href {\doibase
  10.1103/PhysRevB.94.144516} {\bibfield  {journal} {\bibinfo  {journal} {Phys.
  Rev. B}\ }\textbf {\bibinfo {volume} {94}},\ \bibinfo {pages} {144516}
  (\bibinfo {year} {2016})}\BibitemShut {NoStop}%
\bibitem [{\citenamefont {Fu}\ and\ \citenamefont {Kane}(2008)}]{FuL}%
  \BibitemOpen
  \bibfield  {author} {\bibinfo {author} {\bibfnamefont {L.}~\bibnamefont
  {Fu}}\ and\ \bibinfo {author} {\bibfnamefont {C.~L.}\ \bibnamefont {Kane}},\
  }\href {\doibase 10.1103/PhysRevLett.100.096407} {\bibfield  {journal}
  {\bibinfo  {journal} {Phys. Rev. Lett.}\ }\textbf {\bibinfo {volume} {100}},\
  \bibinfo {pages} {096407} (\bibinfo {year} {2008})}\BibitemShut {NoStop}%
\bibitem [{\citenamefont {Sau}\ \emph {et~al.}(2010)\citenamefont {Sau},
  \citenamefont {Lutchyn}, \citenamefont {Tewari},\ and\ \citenamefont
  {Das~Sarma}}]{Sau2010}%
  \BibitemOpen
  \bibfield  {author} {\bibinfo {author} {\bibfnamefont {J.~D.}\ \bibnamefont
  {Sau}}, \bibinfo {author} {\bibfnamefont {R.~M.}\ \bibnamefont {Lutchyn}},
  \bibinfo {author} {\bibfnamefont {S.}~\bibnamefont {Tewari}}, \ and\ \bibinfo
  {author} {\bibfnamefont {S.}~\bibnamefont {Das~Sarma}},\ }\href {\doibase
  10.1103/PhysRevLett.104.040502} {\bibfield  {journal} {\bibinfo  {journal}
  {Phys. Rev. Lett.}\ }\textbf {\bibinfo {volume} {104}},\ \bibinfo {pages}
  {040502} (\bibinfo {year} {2010})}\BibitemShut {NoStop}%
\bibitem [{\citenamefont {Lutchyn}\ \emph {et~al.}(2010)\citenamefont
  {Lutchyn}, \citenamefont {Sau},\ and\ \citenamefont {Das~Sarma}}]{Lutchyn}%
  \BibitemOpen
  \bibfield  {author} {\bibinfo {author} {\bibfnamefont {R.~M.}\ \bibnamefont
  {Lutchyn}}, \bibinfo {author} {\bibfnamefont {J.~D.}\ \bibnamefont {Sau}}, \
  and\ \bibinfo {author} {\bibfnamefont {S.}~\bibnamefont {Das~Sarma}},\ }\href
  {\doibase 10.1103/PhysRevLett.105.077001} {\bibfield  {journal} {\bibinfo
  {journal} {Phys. Rev. Lett.}\ }\textbf {\bibinfo {volume} {105}},\ \bibinfo
  {pages} {077001} (\bibinfo {year} {2010})}\BibitemShut {NoStop}%
\bibitem [{\citenamefont {Oreg}\ \emph {et~al.}(2010)\citenamefont {Oreg},
  \citenamefont {Refael},\ and\ \citenamefont {von Oppen}}]{Oreg}%
  \BibitemOpen
  \bibfield  {author} {\bibinfo {author} {\bibfnamefont {Y.}~\bibnamefont
  {Oreg}}, \bibinfo {author} {\bibfnamefont {G.}~\bibnamefont {Refael}}, \ and\
  \bibinfo {author} {\bibfnamefont {F.}~\bibnamefont {von Oppen}},\ }\href
  {\doibase 10.1103/PhysRevLett.105.177002} {\bibfield  {journal} {\bibinfo
  {journal} {Phys. Rev. Lett.}\ }\textbf {\bibinfo {volume} {105}},\ \bibinfo
  {pages} {177002} (\bibinfo {year} {2010})}\BibitemShut {NoStop}%
\bibitem [{\citenamefont {Sato}\ and\ \citenamefont
  {Fujimoto}(2010)}]{Sato2010}%
  \BibitemOpen
  \bibfield  {author} {\bibinfo {author} {\bibfnamefont {M.}~\bibnamefont
  {Sato}}\ and\ \bibinfo {author} {\bibfnamefont {S.}~\bibnamefont
  {Fujimoto}},\ }\href {\doibase 10.1103/PhysRevLett.105.217001} {\bibfield
  {journal} {\bibinfo  {journal} {Phys. Rev. Lett.}\ }\textbf {\bibinfo
  {volume} {105}},\ \bibinfo {pages} {217001} (\bibinfo {year}
  {2010})}\BibitemShut {NoStop}%
\bibitem [{\citenamefont {Wong}\ and\ \citenamefont {Law}(2012)}]{Wong2012}%
  \BibitemOpen
  \bibfield  {author} {\bibinfo {author} {\bibfnamefont {C.~L.~M.}\
  \bibnamefont {Wong}}\ and\ \bibinfo {author} {\bibfnamefont {K.~T.}\
  \bibnamefont {Law}},\ }\href {\doibase 10.1103/PhysRevB.86.184516} {\bibfield
   {journal} {\bibinfo  {journal} {Phys. Rev. B}\ }\textbf {\bibinfo {volume}
  {86}},\ \bibinfo {pages} {184516} (\bibinfo {year} {2012})}\BibitemShut
  {NoStop}%
\bibitem [{\citenamefont {Li}\ \emph {et~al.}(2015)\citenamefont {Li},
  \citenamefont {Chan},\ and\ \citenamefont {Yao}}]{Li2015}%
  \BibitemOpen
  \bibfield  {author} {\bibinfo {author} {\bibfnamefont {Z.-X.}\ \bibnamefont
  {Li}}, \bibinfo {author} {\bibfnamefont {C.}~\bibnamefont {Chan}}, \ and\
  \bibinfo {author} {\bibfnamefont {H.}~\bibnamefont {Yao}},\ }\href {\doibase
  10.1103/PhysRevB.91.235143} {\bibfield  {journal} {\bibinfo  {journal} {Phys.
  Rev. B}\ }\textbf {\bibinfo {volume} {91}},\ \bibinfo {pages} {235143}
  (\bibinfo {year} {2015})}\BibitemShut {NoStop}%
\bibitem [{\citenamefont {Gangadharaiah}\ \emph {et~al.}(2011)\citenamefont
  {Gangadharaiah}, \citenamefont {Braunecker}, \citenamefont {Simon},\ and\
  \citenamefont {Loss}}]{Gangadharaiah2011}%
  \BibitemOpen
  \bibfield  {author} {\bibinfo {author} {\bibfnamefont {S.}~\bibnamefont
  {Gangadharaiah}}, \bibinfo {author} {\bibfnamefont {B.}~\bibnamefont
  {Braunecker}}, \bibinfo {author} {\bibfnamefont {P.}~\bibnamefont {Simon}}, \
  and\ \bibinfo {author} {\bibfnamefont {D.}~\bibnamefont {Loss}},\ }\href
  {\doibase 10.1103/PhysRevLett.107.036801} {\bibfield  {journal} {\bibinfo
  {journal} {Phys. Rev. Lett.}\ }\textbf {\bibinfo {volume} {107}},\ \bibinfo
  {pages} {036801} (\bibinfo {year} {2011})}\BibitemShut {NoStop}%
\bibitem [{\citenamefont {Stoudenmire}\ \emph {et~al.}(2011)\citenamefont
  {Stoudenmire}, \citenamefont {Alicea}, \citenamefont {Starykh},\ and\
  \citenamefont {Fisher}}]{Stoudenmire2011}%
  \BibitemOpen
  \bibfield  {author} {\bibinfo {author} {\bibfnamefont {E.~M.}\ \bibnamefont
  {Stoudenmire}}, \bibinfo {author} {\bibfnamefont {J.}~\bibnamefont {Alicea}},
  \bibinfo {author} {\bibfnamefont {O.~A.}\ \bibnamefont {Starykh}}, \ and\
  \bibinfo {author} {\bibfnamefont {M.~P.~A.}\ \bibnamefont {Fisher}},\ }\href
  {\doibase 10.1103/PhysRevB.84.014503} {\bibfield  {journal} {\bibinfo
  {journal} {Phys. Rev. B}\ }\textbf {\bibinfo {volume} {84}},\ \bibinfo
  {pages} {014503} (\bibinfo {year} {2011})}\BibitemShut {NoStop}%
\bibitem [{\citenamefont {Fidkowski}\ and\ \citenamefont
  {Kitaev}(2010)}]{Fidkowski2010}%
  \BibitemOpen
  \bibfield  {author} {\bibinfo {author} {\bibfnamefont {L.}~\bibnamefont
  {Fidkowski}}\ and\ \bibinfo {author} {\bibfnamefont {A.}~\bibnamefont
  {Kitaev}},\ }\href {\doibase 10.1103/PhysRevB.81.134509} {\bibfield
  {journal} {\bibinfo  {journal} {Phys. Rev. B}\ }\textbf {\bibinfo {volume}
  {81}},\ \bibinfo {pages} {134509} (\bibinfo {year} {2010})}\BibitemShut
  {NoStop}%
\bibitem [{\citenamefont {Thomale}\ \emph {et~al.}(2013)\citenamefont
  {Thomale}, \citenamefont {Rachel},\ and\ \citenamefont
  {Schmitteckert}}]{Thomale2013}%
  \BibitemOpen
  \bibfield  {author} {\bibinfo {author} {\bibfnamefont {R.}~\bibnamefont
  {Thomale}}, \bibinfo {author} {\bibfnamefont {S.}~\bibnamefont {Rachel}}, \
  and\ \bibinfo {author} {\bibfnamefont {P.}~\bibnamefont {Schmitteckert}},\
  }\href {\doibase 10.1103/PhysRevB.88.161103} {\bibfield  {journal} {\bibinfo
  {journal} {Phys. Rev. B}\ }\textbf {\bibinfo {volume} {88}},\ \bibinfo
  {pages} {161103(R)} (\bibinfo {year} {2013})}\BibitemShut {NoStop}%
\bibitem [{\citenamefont {Katsura}\ \emph {et~al.}(2015)\citenamefont
  {Katsura}, \citenamefont {Schuricht},\ and\ \citenamefont
  {Takahashi}}]{Katsura2015}%
  \BibitemOpen
  \bibfield  {author} {\bibinfo {author} {\bibfnamefont {H.}~\bibnamefont
  {Katsura}}, \bibinfo {author} {\bibfnamefont {D.}~\bibnamefont {Schuricht}},
  \ and\ \bibinfo {author} {\bibfnamefont {M.}~\bibnamefont {Takahashi}},\
  }\href {\doibase 10.1103/PhysRevB.92.115137} {\bibfield  {journal} {\bibinfo
  {journal} {Phys. Rev. B}\ }\textbf {\bibinfo {volume} {92}},\ \bibinfo
  {pages} {115137} (\bibinfo {year} {2015})}\BibitemShut {NoStop}%
\bibitem [{\citenamefont {Miao}\ \emph {et~al.}(2017)\citenamefont {Miao},
  \citenamefont {Jin}, \citenamefont {Zhang},\ and\ \citenamefont
  {Zhou}}]{Miao2017}%
  \BibitemOpen
  \bibfield  {author} {\bibinfo {author} {\bibfnamefont {J.~J.}\ \bibnamefont
  {Miao}}, \bibinfo {author} {\bibfnamefont {H.~K.}\ \bibnamefont {Jin}},
  \bibinfo {author} {\bibfnamefont {F.~C.}\ \bibnamefont {Zhang}}, \ and\
  \bibinfo {author} {\bibfnamefont {Y.}~\bibnamefont {Zhou}},\ }\href {\doibase
  10.1103/PhysRevLett.118.267701} {\bibfield  {journal} {\bibinfo  {journal}
  {Phys. Rev. Lett.}\ }\textbf {\bibinfo {volume} {118}},\ \bibinfo {pages}
  {267701} (\bibinfo {year} {2017})}\BibitemShut {NoStop}%
\bibitem [{\citenamefont {Raghu}\ \emph {et~al.}(2008)\citenamefont {Raghu},
  \citenamefont {Qi}, \citenamefont {Honerkamp},\ and\ \citenamefont
  {Zhang}}]{Raghu2008}%
  \BibitemOpen
  \bibfield  {author} {\bibinfo {author} {\bibfnamefont {S.}~\bibnamefont
  {Raghu}}, \bibinfo {author} {\bibfnamefont {X.~L.}\ \bibnamefont {Qi}},
  \bibinfo {author} {\bibfnamefont {C.}~\bibnamefont {Honerkamp}}, \ and\
  \bibinfo {author} {\bibfnamefont {S.~C.}\ \bibnamefont {Zhang}},\ }\href
  {\doibase 10.1103/PhysRevLett.100.156401} {\bibfield  {journal} {\bibinfo
  {journal} {Phys. Rev. Lett.}\ }\textbf {\bibinfo {volume} {100}},\ \bibinfo
  {pages} {156401} (\bibinfo {year} {2008})}\BibitemShut {NoStop}%
\bibitem [{\citenamefont {Hohenadler}\ \emph {et~al.}(2012)\citenamefont
  {Hohenadler}, \citenamefont {Meng}, \citenamefont {Lang}, \citenamefont
  {Wessel}, \citenamefont {Muramatsu},\ and\ \citenamefont
  {Assaad}}]{Hohenadler2012}%
  \BibitemOpen
  \bibfield  {author} {\bibinfo {author} {\bibfnamefont {M.}~\bibnamefont
  {Hohenadler}}, \bibinfo {author} {\bibfnamefont {Z.~Y.}\ \bibnamefont
  {Meng}}, \bibinfo {author} {\bibfnamefont {T.~C.}\ \bibnamefont {Lang}},
  \bibinfo {author} {\bibfnamefont {S.}~\bibnamefont {Wessel}}, \bibinfo
  {author} {\bibfnamefont {A.}~\bibnamefont {Muramatsu}}, \ and\ \bibinfo
  {author} {\bibfnamefont {F.~F.}\ \bibnamefont {Assaad}},\ }\href {\doibase
  10.1103/PhysRevB.85.115132} {\bibfield  {journal} {\bibinfo  {journal} {Phys.
  Rev. B}\ }\textbf {\bibinfo {volume} {85}},\ \bibinfo {pages} {115132}
  (\bibinfo {year} {2012})}\BibitemShut {NoStop}%
\bibitem [{\citenamefont {Morimoto}\ and\ \citenamefont
  {Nagaosa}(2016)}]{Morimoto2016}%
  \BibitemOpen
  \bibfield  {author} {\bibinfo {author} {\bibfnamefont {T.}~\bibnamefont
  {Morimoto}}\ and\ \bibinfo {author} {\bibfnamefont {N.}~\bibnamefont
  {Nagaosa}},\ }\href {\doibase 10.1038/srep19853} {\bibfield  {journal}
  {\bibinfo  {journal} {Sci. Rep.}\ }\textbf {\bibinfo {volume} {6}},\ \bibinfo
  {pages} {19853} (\bibinfo {year} {2016})}\BibitemShut {NoStop}%
\bibitem [{\citenamefont {Rachel}(2018)}]{Rachel_2018}%
  \BibitemOpen
  \bibfield  {author} {\bibinfo {author} {\bibfnamefont {S.}~\bibnamefont
  {Rachel}},\ }\href {\doibase 10.1088/1361-6633/aad6a6} {\bibfield  {journal}
  {\bibinfo  {journal} {Rep. Prog. Phys.}\ }\textbf {\bibinfo {volume} {81}},\
  \bibinfo {pages} {116501} (\bibinfo {year} {2018})}\BibitemShut {NoStop}%
\bibitem [{\citenamefont {Kudo}\ \emph {et~al.}(2019)\citenamefont {Kudo},
  \citenamefont {Yoshida},\ and\ \citenamefont {Hatsugai}}]{Kubo2019}%
  \BibitemOpen
  \bibfield  {author} {\bibinfo {author} {\bibfnamefont {K.}~\bibnamefont
  {Kudo}}, \bibinfo {author} {\bibfnamefont {T.}~\bibnamefont {Yoshida}}, \
  and\ \bibinfo {author} {\bibfnamefont {Y.}~\bibnamefont {Hatsugai}},\ }\href
  {\doibase 10.1103/PhysRevLett.123.196402} {\bibfield  {journal} {\bibinfo
  {journal} {Phys. Rev. Lett.}\ }\textbf {\bibinfo {volume} {123}},\ \bibinfo
  {pages} {196402} (\bibinfo {year} {2019})}\BibitemShut {NoStop}%
\bibitem [{\citenamefont {Pesin}\ and\ \citenamefont
  {Balents}(2010)}]{Pesin2010}%
  \BibitemOpen
  \bibfield  {author} {\bibinfo {author} {\bibfnamefont {D.~A.}\ \bibnamefont
  {Pesin}}\ and\ \bibinfo {author} {\bibfnamefont {L.}~\bibnamefont
  {Balents}},\ }\href@noop {} {\bibfield  {journal} {\bibinfo  {journal} {Nat.
  Phys.}\ }\textbf {\bibinfo {volume} {6}},\ \bibinfo {pages} {376} (\bibinfo
  {year} {2010})}\BibitemShut {NoStop}%
\bibitem [{\citenamefont {Rachel}\ and\ \citenamefont
  {Le~Hur}(2010)}]{Rachel2010}%
  \BibitemOpen
  \bibfield  {author} {\bibinfo {author} {\bibfnamefont {S.}~\bibnamefont
  {Rachel}}\ and\ \bibinfo {author} {\bibfnamefont {K.}~\bibnamefont
  {Le~Hur}},\ }\href {\doibase 10.1103/PhysRevB.82.075106} {\bibfield
  {journal} {\bibinfo  {journal} {Phys. Rev. B}\ }\textbf {\bibinfo {volume}
  {82}},\ \bibinfo {pages} {075106} (\bibinfo {year} {2010})}\BibitemShut
  {NoStop}%
\bibitem [{\citenamefont {Wang}\ and\ \citenamefont
  {Senthil}(2016)}]{Wang2016}%
  \BibitemOpen
  \bibfield  {author} {\bibinfo {author} {\bibfnamefont {C.}~\bibnamefont
  {Wang}}\ and\ \bibinfo {author} {\bibfnamefont {T.}~\bibnamefont {Senthil}},\
  }\href {\doibase 10.1103/PhysRevX.6.011034} {\bibfield  {journal} {\bibinfo
  {journal} {Phys. Rev. X}\ }\textbf {\bibinfo {volume} {6}},\ \bibinfo {pages}
  {011034} (\bibinfo {year} {2016})}\BibitemShut {NoStop}%
\bibitem [{\citenamefont {Maciejko}\ and\ \citenamefont
  {Fiete}(2015)}]{Maciejko2015}%
  \BibitemOpen
  \bibfield  {author} {\bibinfo {author} {\bibfnamefont {J.}~\bibnamefont
  {Maciejko}}\ and\ \bibinfo {author} {\bibfnamefont {G.~A.}\ \bibnamefont
  {Fiete}},\ }\href {\doibase 10.1038/nphys3311} {\bibfield  {journal}
  {\bibinfo  {journal} {Nat. Phys.}\ }\textbf {\bibinfo {volume} {11}},\
  \bibinfo {pages} {385} (\bibinfo {year} {2015})}\BibitemShut {NoStop}%
\bibitem [{\citenamefont {Hatsugai}\ and\ \citenamefont {Kohmoto}(1992)}]{HK}%
  \BibitemOpen
  \bibfield  {author} {\bibinfo {author} {\bibfnamefont {Y.}~\bibnamefont
  {Hatsugai}}\ and\ \bibinfo {author} {\bibfnamefont {M.}~\bibnamefont
  {Kohmoto}},\ }\href@noop {} {\bibfield  {journal} {\bibinfo  {journal} {J.
  Phys. Soc. Jpn.}\ }\textbf {\bibinfo {volume} {61}},\ \bibinfo {pages} {2056}
  (\bibinfo {year} {1992})}\BibitemShut {NoStop}%
\bibitem [{\citenamefont {Phillips}\ \emph {et~al.}(2020)\citenamefont
  {Phillips}, \citenamefont {Yeo},\ and\ \citenamefont {Huang}}]{PYH}%
  \BibitemOpen
  \bibfield  {author} {\bibinfo {author} {\bibfnamefont {P.~W.}\ \bibnamefont
  {Phillips}}, \bibinfo {author} {\bibfnamefont {L.}~\bibnamefont {Yeo}}, \
  and\ \bibinfo {author} {\bibfnamefont {E.~W.}\ \bibnamefont {Huang}},\ }\href
  {\doibase 10.1038/s41567-020-0988-4} {\bibfield  {journal} {\bibinfo
  {journal} {Nat. Phys.}\ } (\bibinfo {year} {2020}),\
  10.1038/s41567-020-0988-4}\BibitemShut {NoStop}%
\bibitem [{\citenamefont {Nickalls}(2006)}]{viete}%
  \BibitemOpen
  \bibfield  {author} {\bibinfo {author} {\bibfnamefont {R.~W.~D.}\
  \bibnamefont {Nickalls}},\ }\href {\doibase 10.1017/S0025557200179598}
  {\bibfield  {journal} {\bibinfo  {journal} {The Mathematical Gazette}\
  }\textbf {\bibinfo {volume} {90}},\ \bibinfo {pages} {203–208} (\bibinfo
  {year} {2006})}\BibitemShut {NoStop}%
\bibitem [{\citenamefont {Shiozaki}\ \emph {et~al.}(2016)\citenamefont
  {Shiozaki}, \citenamefont {Shapourian},\ and\ \citenamefont
  {Ryu}}]{Shiozaki2016}%
  \BibitemOpen
  \bibfield  {author} {\bibinfo {author} {\bibfnamefont {K.}~\bibnamefont
  {Shiozaki}}, \bibinfo {author} {\bibfnamefont {H.}~\bibnamefont
  {Shapourian}}, \ and\ \bibinfo {author} {\bibfnamefont {S.}~\bibnamefont
  {Ryu}},\ }\href {\doibase 10.1103/PhysRevB.95.205139} {\bibfield  {journal}
  {\bibinfo  {journal} {Phys. Rev. B}\ }\textbf {\bibinfo {volume} {95}},\
  \bibinfo {pages} {205139} (\bibinfo {year} {2016})}\BibitemShut {NoStop}%
\end{thebibliography}%

\end{document}